\documentclass[11pt]{article}

\usepackage{amsfonts,amsmath,amssymb,bm,graphicx,hyperref,a4wide,cite,mathabx,subfigure}

\begin{document}

\title{\textbf{Generation of strong magnetic fields in a nascent neutron star accounting for the chiral magnetic effect}}

\author{$^{a,b}$\,Maxim Dvornikov\thanks{maxdvo@izmiran.ru}
\and
$^{a}$\,V.~B.~Semikoz\thanks{semikoz@yandex.ru}
\and
$^{c}$\,D.~D.~Sokoloff\thanks{sokoloff.dd@gmail.com}
\\
\small{$^{a}$\,Pushkov Institute of Terrestrial Magnetism, Ionosphere} \\
\small{and Radiowave Propagation (IZMIRAN),} \\
\small{108840 Troitsk, Moscow, Russia;} \\
\small{$^{b}$\,Physics Faculty, National Research Tomsk State University,} \\
\small{36 Lenin Avenue, 634050 Tomsk, Russia;} \\
\small{$^{c}$\,Faculty of Physics, M.~V.~Lomonosov Moscow State University,} \\
\small{119991 Moscow, Russia}}

\date{}

\maketitle

\begin{abstract}
We propose the mean field dynamo model for the generation of strongest magnetic fields, $B\sim 10^{15}\,{\rm G}$, in a neutron star (NS) accounting for the chiral magnetic effect (CME) driven by the shock in a supernova (SN) progenitor of that NS. The temperature jump at a narrow shock front, where an initial magnetic field existing in inflowing matter rises sharply, is the source of the CME that prevails significantly the erasure of the CME due to the spin-flip through Coulomb collisions in plasma. The growth of the magnetic field just behind the shock given by the instability term $\nabla\times (\alpha {\bf B})$ in induction equation, stops after a successful SN explosion that throws out the mantle of a protoneutron star. As a result, such an explosion interrupts the transfer of strongly magnetized plasma from the shock onto NS surface and leads to the saturation of the  magnetic field. Assuming the rigid protostar rotation, we employ the mean field dynamo, which is similar to the $\alpha^2$-dynamo known in the standard magnetohydrodynamics (MHD).
The novelty of our model is that $\alpha^2$-dynamo is based on concepts of particle physics, applied in MHD, rather than by a mirror asymmetry of convective vortices in the rotating convection.
\end{abstract}

\maketitle

\section{Introduction}

The generation of strongest magnetic fields observed in magnetars $\sim 10^{15}\,{\rm G}$ \cite{KasBel17} is still an open problem in astrophysics. The observation of X-rays and the soft $\gamma$-emission from such objects means the penetration of these fields the magnetosphere of a neutron star (NS), filled by the magnetized plasma, which is a source of electromagnetic waves. The purpose of our study is to understand how such superstrong magnetic fields, with $B\gg B_\mathrm{Schwinger}=m_e^2/e=4.41\times 10^{13}\,{\rm G}$, can arise inside a nascent neutron star (NS).

Rough estimates of the magnetic field strength, generated by the $(\alpha-\Omega)$-dynamo at the early stage of a nascent NS, are given in the pioneering work~\cite{Thompson:1993hn} on this subject. Magnetic fields as strong as $3\times 10^{15}\,{\rm G}$ can be generated as a random superposition of many small dipoles with the size $l\sim 1\,{\rm km}$ through convection instability in such nascent NS originated by a huge neutrino emission in the protoneutron star (PNS) during first seconds after the core bounce. The $\alpha$-helicity parameter for the NS rotation period $P_\mathrm{rot}\sim 30\,\text{ms}$, estimated in Ref.~\cite{Thompson:1993hn} within the $(\alpha-\Omega)$-dynamo model, is $\alpha\sim \Omega l/c\sim 10^{-3}$, where $\Omega=2\pi/P_\mathrm{rot}$ is the angular velocity and $l\simeq 1\,\mathrm{km}$ is the coherence length of the magnetic field. Here we use the dimensionless units $\hbar=c=1$ throughout the text.

The magnetic field origin in various celestial bodies is mainly associated with a dynamo action, i.e. the transformation of the kinetic energy into the magnetic one in an electrically conducting media (see, e.g., Refs.~\cite{Parker2,Zeldovich}). The drivers of dynamo action are thought to be the differential rotation and various forms of mirror asymmetric flows in convective or turbulent shells of the body of interest. The dynamo action in NSs was considered, in particular,  in Ref.~\cite{Thompson:1993hn} (see also Ref.~\cite{Bonnano}) for PNS in the context of the pulsar magnetic field origin.
 
However, the point is that a pulsar magnetic field can be thought to be amplified from a relatively weak magnetic field in a normal star during its compression into a neutron star. Perhaps, it is why the NS dynamo does not attract the main attention of experts in dynamo studies, who more prefer solar or galactic dynamos. One more point is that NSs {\it as remnants of a supernova (SN) explosion} are believed to be solid body rotators (see, e.g., Sec.~6.12 in Ref.~\cite{Yakovlev}), while the differential rotation is a very popular (however, not obligatory) dynamo driver for dynamo modelers. 

According to the contemporary knowledge, dynamo driven stellar magnetic fields can reach the equipartition between magnetic and kinetic energies. It seems to be sufficient for the explanation of pulsar magnetic fields even if details of the magnetic field evolution from a relatively weak magnetic field in a normal star to that in NS remain debatable. The problem arises for superstrong magnetar magnetic fields, where an additional amplification in NS is likely to be inevitable even if various above discussed options are exploited.

Here, we face a problem with the concept of the solid body rotation of PNS since it is much easier to amplify quasi-stationary magnetic fields using the joint action of the differential rotation and the mirror asymmetry rather by a mirror asymmetry only. Nevertheless, a dynamo based on the mirror asymmetry only, known as $\alpha^2$-dynamo, is still possible and widely discussed in stellar astrophysics (see, e.g., Ref.~\cite{CK06}). The novelty of our suggestion is that the $\alpha^2$-dynamo is based on some concepts in particle physics implemented for the Anomalous Magneto-HydroDynamics (AMHD) rather owing to the mirror asymmetry of convective vortexes in the rotating convection.

Some new attempts to explain the generation of strong magnetic fields in PNS using AMHD were undertaken, e.g., in Refs.~\cite{SigLei16,MasKotYam18,Grabowska:2014efa}. The following development of AMHD as laminar and turbulent dynamos applying such approaches both for a hot plasma in the early Universe and for the degenerate ultrarelativistic electron plasma in PNS was made in Refs.~\cite{Rogachevskii:2017uyc,Schober:2017cdw}. Note that, in Ref.~\cite{Schober:2018wlo}, the mean spin term in a hot plasma, $\sim {\bf B}\nabla\mu_e$, was taken into account as an additional source for the evolution of a chiral anomaly in AMHD. However, in a hot plasma of the early universe, the electron chemical potential $\mu_e$ is negligible, $\mu_e\ll T$. Therefore the lepton asymmetry and the corresponding mean spin cease.

In the present work, we continue our studies for the application of AMHD trying to explain the generation of strong magnetic fields in PNS. The mirror asymmetric magnetic field driver, responsible for the dynamo action, is connected with the mirror asymmetry of some particle physics characteristics.\footnote{In AMHD, the $\alpha$-helicity parameter is proportional to the chiral imbalance $\mu_5=(\mu_\mathrm{R} - \mu_\mathrm{L})/2$, where $\mu_\mathrm{R,L}$ are the chemical potentials of right and left-handed electrons and $\alpha\sim \mu_5$. Such a parameter is preudoscalar, i.e. it is mirror asymmetric, since $\mu_5 \to - \mu_5$ under the space inversion.} In this sense, accounting for the rigid rotation $\Omega=\text{const}$ assumed in our work, the dynamo suggested here can be considered as a specific kind of the so-called $\alpha^2$-dynamo based on the mirror asymmetry of convective or turbulent flows in the standard MHD (see, e.g., Ref.~\cite{Krause}). 

A substantial difference of the process under discussion from the conventional forms of $\alpha^2$-dynamo is that the dynamo driver in question is connected with a magnetic field somehow created at the previous stage of the magnetar evolution, namely in the SN progenitor of a nascent NS as we suggest here. Correspondingly, the process occurs to be very nonlinear just from its starting point. It is why the application of conventional methods of dynamo studies to separate the investigation of the evolution of the dynamo driven magnetic field from the evolution of dynamo drivers in order to simplify the problem, occurs here even more limited rather in stellar or galactic dynamos.

Of course, contemporary numerical simulations can be applied for the solution of very complicated equations, and it is not a great deal, in principle, to include various effects in a single numerical model. Typically, it would require to consider various details of the dynamo driven magnetic field configuration. The point however is that our knowledge of the structure and physical properties of magnetars is very limited. Thus, it would be highly desirable to avoid detailed description of the magnetar magnetic field configuration and deal with few of its first Fourier modes.

In order to resolve these contradictory intentions, we consider a general set of governing equations in a suitable Fourier basis and truncate it as strongly as possible to keep a few Fourier modes to allow the dynamo action. A simple analysis, e.g., in Ref.~\cite{Sokoloff}, shows that we need here two Fourier modes for the poloidal magnetic field component and two Fourier modes for the toroidal magnetic field component. Correspondingly, we keep two Fourier modes for the $\alpha$-effect as well. Of course, our dynamo model is quite sketchy and illustrative. However, we believe that it allows to understand the physics of the problem under discussion until one learns more about the magnetar structure.       

In short, the scenario for generation of strong magnetic fields is the following. We consider the amplification of an initial magnetic field within the envelope of the contracting protostar at the stage after shock bounce in the PNS core, and rely here on the model for a revival of the stalled supernova shock previously suggested in Ref.~\cite{Janka}. Note that the recent 3D calculations confirm a delay of the stalled shock at $R_s\simeq 150\,\text{km}$ during a long time $t_\mathrm{pb}\sim 0.3\,\text{s}$ (see Fig.~1 in Ref.~\cite{Melson:2019kjj}). Since the gas falling into the stalled shock is strongly decelerated and postshock velocities are smaller than the local sound speed, the structure of collapsed star, including the region behind the shock, can be well described by hydrostatic equilibrium~\cite{Janka2}.   Therefore, in our AMHD dynamo model, we consider all parameters, such as the conductivity, the density, the temperature, in the radiation dominated region just behind the shock as the static ones. The evolution of the magnetic field components and the $\alpha$-helicity parameter proceeds there much faster. Hence we solve self-consistently the full set of the mean field dynamo equations as a problem with chosen initial conditions.

Thus, in the present work we develop the mean field dynamo approach in AMHD for the generation of strong magnetic fields in NS just after the core bounce in its SN progenitor. The amplification of an initial magnetic field is based on the chiral magnetic effect (CME) in relativistic electron gas. The CME is effective both within the shock front and just behind shock crossed by inflowing mantle, which, in turn, transfers such magnetic field frozen in plasma and amplified at the stalled shock position $R_s\simeq 150\,\mathrm{km}$ to the NS surface at $R_\mathrm{NS}=10\,\mathrm{km}$. As a result, the additional growth of the magnetic field frozen in plasma on more than two orders of magnitude can occur owing to that transfer.   

Our work is organized as follows. In Sec.~\ref{scenario}, we describe, in short, the scenario for the magnetic field amplification near the shock in SN. Then, in Sec.~\ref{sec:CHIRANOMSN}, we derive the master evolution equation for the $\alpha$-helicity parameter that enters the magnetic field instability term in the Faraday  equation, $\sim \nabla\times \alpha {\bf B}$. For that we start from the statistically averaged axial Ward anomaly, known as the Adler-Bell-Jackiw anomaly in QED, estimating, in Sec.~\ref{sec:CONDSHOCK}, the static electric conductivity $\sigma$ behind shock. In Sec.~\ref{sec:ALPHASHOCK}, we present the master evolution equation for $\alpha (t)$ in the dimensionless form suitable for numerical simulations. The full set of the evolution equations in the mean field AMHD dynamo approach is given in Sec.~\ref{dynamo}. In Sec.~\ref{low}, we formulate the low Fourier mode approximation and, in Sec.~\ref{sec:inicond}, we choose the initial conditions. Then, in Sec.~\ref{sec:RESULTS}, we present the results of the numerical simulations showing the dynamo amplification of an initial magnetic field and the evolution of the $\alpha$-helicity parameter for the different shock widths $(\Delta r)_\mathrm{front}$. We discuss our results in Sec.~\ref{sec:DISC}. In Appendix~\ref{add}, we present the full set of the ordinary non-linear differential equations solved numerically in our model.

\section{The CME scenario for the magnetic field generation by the shock in SN\label{scenario}}

 In Fig.~2 in Ref.~\cite{Janka}, one can see the heating region behind the shock where, first, there is a jump of the temperature and the density at the shock position $r=R_s$. Note also that the electron number density for a mildly degenerate electron gas in that hot region, $\mu_e >T > m_e$, see Eq.~(57) in Ref.~\cite{Janka},
\begin{equation}\label{edensity1}
n_e=Y_e\frac{\rho}{m_N}=\frac{\mu_e^3}{3\pi^2} + \frac{T^2\mu_e}{3}, 
\end{equation}
should be valid for $T\ll \mu_e$ and differs from the case of fully degenerate gas because of the temperature correction.

When a shock passes the neutrinosphere at the distance $r=R_{\nu}=50\,\mathrm{km}> R_\mathrm{NS}=10\,\mathrm{km}$, one finds from Eq.~(\ref{edensity1}), accounting for the temperature $T_{\nu}=4\,{\rm MeV}$ and the corresponding density $n_e$,\footnote{The electron density at the neutrinosphere, $n_e=(1/\sigma_w R_{\nu})=10^{36}\,\text{cm}^{-3}$, is given by the cross section of the electroweak $\nu e$-scattering, $\sigma_w=2\times 10^{-44}(E_{\nu}/{\rm MeV})\,\text{cm}^2$, where $E_{\nu}\simeq 10\,{\rm MeV}$ is the mean electron neutrino energy emitted from SN.} the ultrarelativistic chemical potential $\mu_e=42\,{\rm MeV}\gg m_e$. While for a stalled shock at the distance $r=R_s=150\,\mathrm{km}$, using estimates for the density of matter consisting of electrons, protons, and neutrons, $\rho\simeq 10^9\,\text{g}\cdot\text{cm}^{-3}$, $Y_e=0.2$, and $n_e=2\times 10^{32}\,\text{cm}^{-3}$, as well as the temperature $T_s\simeq 1\,{\rm MeV}$ behind the shock~\cite{Janka}, we find the case of the mildly degenerate electron gas, $\mu_e=3\,{\rm MeV}> T_s>m_e$.

On the other hand, it is well known that, due to the direct Urca process in a contracting protostar  with emission of left neutrinos, $p + e_\mathrm{L}\to n + \nu_{e\mathrm{L}}$, there appears in the environment before the core bounce  a small imbalance between densities of right-handed and left-handed electrons, $n_5=n_\mathrm{R} - n_\mathrm{L}>0$, $n_5\ll n_e$, where $n_e=n_\mathrm{R} + n_\mathrm{L}$,
\begin{equation}\label{n5new1}
n_5(\mathbf{x},t)=\left[\frac{\mu_\mathrm{R}^3}{6\pi^2} - \frac{\mu_\mathrm{L}^3}{6\pi^2}\right] + \frac{T^2}{3}\Bigl(\mu_\mathrm{R} - \mu_\mathrm{L}\Bigr)=\mu_5(t)\frac{\mu_e^2}{\pi^2}\left[1 + \frac{2\pi^2}{3}\left(\frac{T}{\mu_e}\right)^2\right].
\end{equation}
The evolution of the chemical potential imbalance $\mu_5(t)=(\mu_\mathrm{R} - \mu_\mathrm{L})/2$ that enters the anomalous  current ${\bf j}_\mathrm{anom}= e^2\mu_5{\bf B}/2\pi^2$ in the Maxwell equation, $\nabla\times {\bf B}={\bf j} + {\bf j}_\mathrm{anom}$,\footnote{The presence of such electric current is called in literature as the chiral magnetic effect (CME) in relativistic plasma~\cite{Vilenkin:1980fu,Kharzeev:2015znc}. Note that ${\bf j}_\mathrm{anom}$ is the vector current, since $\mu_5$ is pseudoscalar under the space inversion, for which $n_\mathrm{R}\leftrightarrow n_\mathrm{L}$, or $\mu_5\to - \mu_5$.} added to the ohmic current ${\bf j}=\sigma ({\bf E} + {\bf v}\times {\bf B})$, is driven by a huge temperature gradient at the narrow shock front $\mathrm{d}T/\mathrm{d}r|_{r=R_s}\simeq -\Delta T/(\Delta r)_\mathrm{front}$, see Eq.~(\ref{mu5}) below. Such a temperature gradient is seen at the distance $r=R_s$ from the NS center in Fig.~2 in Ref.~\cite{Janka}.

\subsection{Shock width in supernova}

Here a narrow shock width, $(\Delta r)_\mathrm{front}$, in general, should be of order the free path\footnote{For an intensive shock its width should be of order the free path $\lambda$ for particles in the corresponding medium~\cite{LL}. For instance, in the terrestrial atmosphere the shock width $(\Delta r)_\mathrm{front}\sim \lambda= 1/n_\mathrm{L}\sigma\simeq 10^{-4}\,\text{cm}$ is given by the molecule density (Loschmidt number), $n_\mathrm{L}=2.7\times 10^{19}\,\text{cm}^{-3}$, and the cross-section for molecule collisions, $\sigma\simeq \pi a^2$, where $a=10^{-8}\,\text{cm}$.} for nuclei coming from protostar matter inflow  with respect to their photo-disintegration into free nucleons just within the shock front at $R_s=150\,\mathrm{km}$. It means that the width is large enough, $(\Delta r)_\mathrm{front} \sim \lambda_{\gamma}=(\sigma_{\gamma}n_\mathrm{B})^{-1}\sim 10^{-8}\,\text{cm}$. Here $n_\mathrm{B}=n_e/Y_e\simeq 10^{33}\,\text{cm}^{-3}$ is the baryon density for the electron density in Eq.~(\ref{edensity1}) behind the shock front and $\sigma_{\gamma}=(20\div 30)\,{\rm mb}$ is the cross section accounting for the giant dipole resonance for the photo-disintegration reactions \cite{Ishkhanov}.

Before that time, at an earlier moment, the intensive shock passes the neutrinosphere at $R_{\nu}=50\,\mathrm{km}$, where the baryon density is greater than for a stalled shock at $R_s=150\,\mathrm{km}$, $n_\mathrm{B}=n_e/Y_e=5\times 10^{36}\,\text{cm}^{-3}$. As a result, the shock width is much shorter there, $(\Delta r)_\mathrm{front}=(\sigma_{\gamma}n_\mathrm{B})^{-1}=10^{-11}\,\text{cm}$. This fact strongly affects the CME growth of the $\alpha$-helicity parameter in the mean field  dynamo. However, at that time, we do not expect the successful SN explosion, which leads to the shutdown of magnetized protostar matter inflow, hence, to a fast saturation of the CME amplification of a seed magnetic field.

Therefore, to shorten characteristic times for the AMHD dynamo, we assume below a finite shock width at $R_s=150\,\mathrm{km}$ which is less than $\lambda_{\gamma}$ above, $(\Delta r)_\mathrm{front} \sim (10^{-11}\div 10^{-10})\,\text{cm}$, where the shortest one corresponds to the mean distance between particles just behind shock $r\lesssim R_s$, $(\Delta r)_\mathrm{front}\simeq n_\mathrm{B}^{-1/3}\sim 10^{-11}\,\text{cm}$. This assumption is a compromise between the two contradictive demands: the free path in gas-dynamics of continuous media should be zero, $\lambda=0$, or gas-dynamical methods fail for the description of the inner structure of intensive shocks~\cite{LL}. While, on the other hand, a real shock width should be finite.

\subsection{Attenuation of the magnetic field instability due to successful SN explosion}

Somewhere behind the shock at temperature $T> 1\,{\rm MeV}$ the dense matter consists of free electron-positron pairs, photons and nucleons in the radiation-dominated region where neutrinos (antineutrinos) loose energy through their captures, $\nu_e + n\to e^- + p$ and $\bar{\nu}_e + p\to e^+ + n$. It results in the shock reheating, just changing its possible stagnation regime. For sufficiently large neutrino luminosities $L_{\nu}$ such reheating should lead to the successful explosion throwing out the PNS mantle, as shown in Figs.~6 and~7 in Ref.~\cite{Janka}. The corresponding reversal of the matter inflow to its outflow interrupts the transfer of magnetized plasma from the shock onto NS surface. This stage leads to the end of the magnetic field instability given by the term $\sim \nabla\times \alpha {\bf B}$ in the Faraday induction equation, see below in Eq. (\ref{Faraday1}). Consequently, the helicity parameter $\alpha\sim \mu_5$ of the CME origin driven by the shock in a supernova ceases somehow, $\alpha\to 0$, see here Eq. (\ref{cutoff}) and Fig.~\ref{fig:alphagrowth} .

Since the AMHD dynamo action is presumably faster than the acceleration of shock due to reheating, or such plasma parameters as the temperature, the density, the conductivity are static at $R_s$, we do not touch a more general problem of the common solution of the hydrodynamical equations for a supernova explosion, see corresponding Eqs.~(2)-(6) in Ref.~\cite{Janka}, and AMHD dynamo for magnetic fields.  We introduce instead the single temporal factor $F(t)$ that simulates the reversal of the matter inflow to its outflow in hydrodynamics, leading to the CME cutoff $\alpha\to 0$ via the change $\alpha\to F(t)\alpha$,
\begin{equation}\label{cutoff}
  F(t)=\frac{1}{2}\left\{1 - \tanh [K(t-t_0)]\right\}.
\end{equation}
Here $K=10/t_0$, $t_0$ is the CME cutoff time that varies in dependence on $(\Delta r)_\mathrm{front}$ related with the intensity of the shock\footnote{The shock width reduces with the increase of shock intensity given by a growth of the pressure jump, $(\Delta r)_\mathrm{front}\sim (P_s - P_p)^{-1}$, where $P_{s}$ is the postshock pressure and $P_{p}$ is the preshock one, see Eq.~(93.13) in Ref.~\cite{LL}. The more intensive shock due to a great neutrino luminosity $L_{\nu}$ in PNS, the earlier the successful SN explosion starts, see in Figs.~6 and~7 in Ref.~\cite{Janka}.}, see below in Sec.~\ref{sec:RESULTS}. For late times $t\gg t_0$ the product $F(t)\alpha$ vanishes together with the CME, $F(t)\alpha\to 0$. The instability term $F(t)\nabla\times \alpha {\bf B}$ ceases in the induction Eq.~(\ref{Faraday}) resulting in the saturation of ${\bf B}$ (see Fig.~\ref{fig:fieldcomponents}). Obviously, this factor is not important for $t< t_0$ since $F(t)\approx 1$. Hence, the CME is active only at early times.

\section{Chiral anomaly evolution in the SN progenitor of a nascent NS\label{sec:CHIRANOMSN}}

We consider the axial Ward anomaly,
\begin{equation}\label{pseudovector}
\frac{\partial }{\partial x^{\mu}}\bar{\psi}_e\gamma^{\mu}\gamma^5\psi_e=\frac{\partial j_\mathrm{R}^{\mu}}{\partial x^{\mu}} - \frac{\partial j_\mathrm{L}^{\mu}}{\partial x^{\mu}} = 2\mathrm{i}m_e\bar{\psi}_e\gamma_5\psi_e + \frac{2e^2}{16\pi^2}F_{\mu\nu}\tilde{F}^{\mu\nu},
\end{equation}
statistically averaged in a magnetized degenerate ultrarelativistic electron gas, $p_{F_e}\gg m_e$,
\begin{equation}\label{n5evolution}
\frac{\partial n_5(\mathbf{x},t)}{\partial t} =  - \nabla\cdot \mathbf{S}(\mathbf{x},t)+\frac{2\alpha_\mathrm{em}}{\pi}({\bf E}\cdot\mathbf{B}) -\Gamma_f n_5(\mathbf{x},t),
\end{equation}
where $n_5(\mathbf{x},t)=n_\mathrm{R}(\mathbf{x},t) - n_\mathrm{L}(\mathbf{x},t)=\langle\psi^+_e\gamma_5\psi_e\rangle_0\sim \mu_5$ is the chiral density in Eq.~(\ref{n5new1}), a small mass pseudoscalar term $2\mathrm{i}m_e\langle\bar{\psi}_e\gamma_5\psi_e\rangle_0\sim m_e$ was omitted, and we added the spin-flip term $\sim \Gamma_f$ arising due to Coulomb collisions in plasma with the chirality flip, $e_\mathrm{L}\leftrightarrow e_\mathrm{R}$, that diminishes chiral imbalance, $\mu_5\to 0$ . 
Here the term  $\nabla\cdot{\bf S}$ given by the mean spin ${\bf S}=\langle\psi_e^+\bm{\Sigma}\psi_e\rangle_0= - e\mu_e{\bf B}/2\pi^2$, known as the chiral separation effect (CSE)~\cite{Kharzeev:2015znc}, is fully canceled\footnote{Note that such a cancellation is absent in the early universe plasma since the early universe matter is uniform.} due to the term $\nabla\mu_e/e$ entering the electric field in the generalized Ohm law for a non-uniform medium,\footnote{To get Eq.~(\ref{electric}) we equate the effective electric field ${\bf E}^*$ in Eq.~(19) to another form for ${\bf E}^*$ in Eq.~(26) in Ref.~\cite{Shternin} adding anomalous current ${\bf j}_\mathrm{anom}= e^2\mu_5{\bf B}/2\pi^2$ for chiral  plasma.}
\begin{equation}\label{electric}
{\bf E}= - {\bf v}\times {\bf B} + \frac{{\bf j} + {\bf j}_\mathrm{anom}}{\sigma} - \frac{\nabla\mu_e}{e} - \left[Q_\mathrm{T} + \frac{\mathcal{S}}{en}\right]\nabla T,
\end{equation}
where $\sigma$ is the electric conductivity, $Q_\mathrm{T}$ is the thermopower  from Eq. (46c) in Ref. \cite{Shternin} when substituting there $m_e^*=\mu_e\approx p_{F_e}$,\footnote{We remind that we use units $k_\mathrm{B}=c=\hbar=1$.}
\begin{equation}
  Q_\mathrm{T}=\frac{4\pi^2}{3e}\left(\frac{T}{p_{F_e}}\right),
\end{equation}
In Eq.~\eqref{electric}, $\mathcal{S}$ is the entropy density. The entropy per baryon, $\mathcal{S}/n$, estimated in Ref.~\cite{Thompson:1993hn} for early times after the core bounce is $\mathcal{S}/n\simeq 5\div 10$ . Thus, substituting the mean spin term $-\nabla\cdot{\bf S}=+e{\bf B}\nabla\mu_e/2\pi^2$ and the electric field in Eq.~(\ref{electric}) to Eq.~(\ref{n5evolution}), one gets the cancellation of the mean spin term. We account for the expression of product $\sim {\bf E}\cdot{\bf B}$ in Eq.~(\ref{n5evolution}),
\begin{align}
\frac{2\alpha_\mathrm{em}}{\pi}({\bf E}\cdot{\bf B})=&
\frac{2\alpha_\mathrm{em}}{\pi\sigma}(\nabla\times {\bf B})\cdot{\bf B} -  \frac{(2\alpha_\mathrm{em}^2)}{\pi^2\sigma}\mu_5{\bf B}^2 - \frac{e{\bf B}\nabla \mu_e}{2\pi^2}
\nonumber
\\
&- \frac{e^2}{2\pi^2}\left[\frac{4\pi^2}{3e}\left(\frac{T}{p_{F_e}}\right) + \frac{S}{en}\right]{\bf B}\nabla T,
\end{align} 
where $\alpha_\mathrm{em}=e^2/4\pi\approx (137)^{-1}$ is the fine structure constant.
Thus, accounting for the value of $-\nabla\cdot {\bf S} + (2\alpha_\mathrm{em}/\pi)({\bf E}\cdot{\bf B})$, we get from Eq.~(\ref{n5evolution}) the evolution equation for the chiral density $n_5=n_\mathrm{R} - n_\mathrm{L}$ , where the temperature gradient at  the shock front $\sim \nabla T$ is the source of the CME,
\begin{align}\label{n5new} 
\frac{\partial n_5(\mathbf{x},t)}{\partial t} = &
-\left[\frac{2e}{3}\left(\frac{T}{p_{F_e}}\right) + \frac{e}{2\pi^2}\left(\frac{\mathcal{S}}{n}\right)\right]{\bf B}\nabla T + \frac{2\alpha_\mathrm{em}}{\pi\sigma}[{\bf B}\cdot(\nabla\times {\bf B})]
\nonumber
\\
& -  n_5(\mathbf{x},t)\left(\Gamma_f + \frac{(2\alpha_\mathrm{em})^2}{\sigma\mu_e^2}\left[1 + \frac{2\pi^2}{3}\left(\frac{T}{\mu_e}\right)^2\right]^{-1}{\bf B}^2\right).
\end{align}
Basing on Eq.~\eqref{n5new}, one gets the kinetic equation for the chiral anomaly imbalance $\mu_5$ related with $n_5$ in Eq. (\ref{n5new1}),
\begin{align}\label{mu5}
\frac{\partial \mu_5(\mathbf{x},t)}{\partial t} =&
-\frac{\pi^2}{\mu_e^2}\left[1 + \frac{2\pi^2}{3}\left(\frac{T}{\mu_e}\right)^2\right]^{-1}\left[\frac{2e}{3}\left(\frac{T}{p_{F_e}}\right) +\frac{e}{2\pi^2}\left(\frac{\mathcal{S}}{n}\right)\right]{\bf B}\nabla T
\nonumber
\\
&  - \Gamma_f\mu_5\left[1 + \frac{{\bf B}^2}{B_0^2}\right],
\end{align}
where we omitted a small magnetic diffusion term $\sim {\bf B}\cdot(\nabla\times {\bf B})/\sigma$, and input the normalization, ${\bf B}^2/B_0^2$,
\begin{equation}\label{B0}
B_0^2=\frac{\Gamma_f\sigma\mu_e^2}{(2\alpha_\mathrm{em})^2}\left[1 + \frac{2\pi^2}{3}\left(\frac{T}{\mu_e}\right)^2\right]=1.75\,{\rm MeV}^4, 
\end{equation}
that corresponds to $B_0=1.33\,{\rm MeV}^2=6.6\times 10^{13}\,{\rm G}$.
Here $\sigma=17\,{\rm MeV}$ is the electric conductivity near the shock, see below Eq.~(\ref{conductivity2}), and $\Gamma_f$ is the rate of the spin-flip, $e_\mathrm{R}\leftrightarrow e_\mathrm{L}$, due to Coulomb $ep$ collisions~\cite{Grabowska:2014efa},
\begin{equation}\label{Gammaf}
\Gamma_f=\frac{\alpha_\mathrm{em}^2}{3\pi}\left(\frac{m_e}{\mu_e}\right)^2\mu_e\left[\ln \left(\frac{1}{\alpha_\mathrm{em}(1 + \mu_e/3T)}\right) -1\right],
\end{equation}
where we substitute $\mu_e=3\,{\rm MeV}$ and $T=1\,{\rm MeV}$ resulting in $\Gamma_f=1.43\times 10^{-6}\,{\rm MeV}$.

\subsection{Electric conductivity near the shock\label{sec:CONDSHOCK}}

We consider matter, consisting of electrons $e$, protons $p$, and neutrons $n$, in the radiation dominated region near the shock at $r\lesssim R_s=150\,\mathrm{km}$ just after the core bounce and use the electric conductivity in Eq.~(74) in Ref.~\cite{Shternin},
\begin{equation}\label{conductivity}
\sigma=1.86\times 10^{30}\,\text{s}^{-1}\left(\frac{n_\mathrm{B}}{n_0}\right)^{8/9}\times\left(\frac{T}{10^8\,{\rm K}}\right)^{-5/3}A(x_e,x_p), 
\end{equation}
where we neglect the contribution of muons. To obtain Eq.~\eqref{conductivity} we put $x_e=n_e/n_\mathrm{B}=x_p=n_p/n_\mathrm{B}=Y_e=0.2$ using the electroneutrality condition $n_e=n_p$, $A(x_e,x_p)\sim 1$. Here $n_0=0.16\,\text{fm}^{-3}=1.28\times 10^6\,{\rm MeV}^3$ is the central density within NS saturated core, or $\rho_0=2.7\times 10^{14}\,\text{g}\cdot\text{cm}^{-3}$, and $n_\mathrm{B}$ is the baryon density near the shock, $n_\mathrm{B}\ll n_0$.

Note that, in Ref.~\cite{Janka}, one assumes a rarefied matter $n_\mathrm{B}=n_e/0.2=10\,{\rm MeV}^3$ near the shock position $R_s$, since $\mu_e=3\,{\rm MeV}$ and $T= 1\,{\rm MeV}$. It  results in $n_e= \mu_e^3/3\pi^2 + \mu_eT^2/3\sim 2\,{\rm MeV}^3$. Taking into account that  $n_\mathrm{B}/n_0 = 7.8\times 10^{-6}$, we get that 
\begin{equation}\label{conductivity2}
  \sigma\simeq 17\,{\rm MeV},
\end{equation}
where we use Eq.~\eqref{conductivity}.

\subsection{Evolution of the helicity parameter driven by shock in SN\label{sec:ALPHASHOCK}}

From the statistically averaged chiral anomaly in Eq. (\ref{mu5}) one obtains the master evolution equation for the magnetic helicity parameter $\alpha=(2\alpha_\mathrm{em}\mu_5)/\pi\sigma$ in the form,
\begin{equation}\label{alpha}
\frac{\partial \alpha }{\partial \tau }= -\frac{2\alpha_\mathrm{em}\pi R_s^2}{\mu_e^2}\left[\frac{e(2T/3\mu_e) + e(\mathcal{S}/2\pi^2n)}{1 + 2\pi^2(T/\mu_e)^2/3}\right]B_r\frac{{\rm d}T}{{\rm d}r} - \alpha\left[\Gamma_f\sigma R_s^2\left(1 + \frac{{\bf B}^2}{B_0^2}\right)\right],
\end{equation}
where $\sigma= 17\,{\rm MeV}$ results from Eq.~(\ref{conductivity2}) for $T=1\,{\rm MeV}$, $\mu_e=3\,{\rm MeV}$, then $R_s=1.5\times 10^7\,\text{cm}$ is the shock distance from NS center, $\Gamma_f=1.43\times 10^{-6}\,{\rm MeV}$ is the rate of the spin-flip in Eq.~\ref{Gammaf}), $\tau=t/t_\mathrm{diff}$ is the dimensionless time, and $t_\mathrm{diff}=\sigma R_s^2=2\times 10^8\,\text{yr}$ is the magnetic diffusion time.

To proceed in the analysis of Eq.~\eqref{alpha}, we use the spherical coordinates system, that is natural if one deals with the magnetic field in NS. Then, following Ref.~\cite[p.~373]{Sti04}, we decompose the magnetic field into the toroidal $\mathbf{B}_t = B_\varphi \mathbf{e}_\varphi$ and the poloidal $\mathbf{B}_p = B_r \mathbf{e}_r + B_\theta \mathbf{e}_\theta$ components: $\mathbf{B} = \mathbf{B}_t + \mathbf{B}_p$. Moreover, we introduce the vector potential $\mathbf{A}= A_\varphi \mathbf{e}_\varphi$ for $\mathbf{B}_p = (\nabla \times \mathbf{A})$. 

We can express the poloidal components $B_{r,\theta}$ for the axisymmetric field given by the azimuthal potential $A_{\varphi}$ in the form,
\begin{equation}\label{components}
B_{\theta}=- \frac{1}{r}\frac{\partial}{\partial r}(rA_{\varphi}),
\quad
B_{r}=\frac{1}{r\sin\theta}\frac{\partial}{\partial \theta}(\sin\theta A_{\varphi}),
\end{equation}
where $\theta$ is the colatitude angle. Finally, accounting for the factor $\Gamma_f\sigma R_s^2=1.36\times 10^{31}$ and using Eqs.~\eqref{alpha} and~\eqref{components}, one obtains
\begin{equation}\label{master}
\frac{\partial \alpha }{\partial \tau }= + \frac{1.8\times 10^{22}}{(\Delta r)_\mathrm{front}/1\,\text{cm}}\left(\frac{\partial A}{\partial \theta} + \cot \theta A\right) - \alpha\left[1.36\times 10^{31}\left(1  + \frac{{\bf B}^2}{B_0^2}\right)\right],
\end{equation}
where we substituted $\mathrm{d}T/\mathrm{d}r=-T/(\Delta r)_\mathrm{front}$.

\section{AMHD dynamo model for the rigid protostar rotation\label{dynamo}}

For simplicity we consider the rigid PNS rotation $\Omega=\text{const}$, for which the differential rotation is absent, $\partial_{\theta}\Omega=\partial_r\Omega=0$. Hence the dynamo term $\nabla\times ({\bf v}\times {\bf B})$  vanishes in the Faraday equation that takes the form,
\begin{equation}\label{Faraday1}
\frac{\partial {\bf B}}{\partial t} =  \frac{1}{\sigma}\nabla^2{\bf B} + \nabla\times \alpha {\bf B}.
\end{equation}
Therefore, we should not involve the Navier-Stokes equation neglecting also a small vorticity contribution $\sim \bm{\omega}=\nabla\times {\bf v}$. 

The complete system of the dynamic equations for the 3D dynamo model with an axisymmetric magnetic field in a thin layer at $r\approx R_s$ includes the following three AMHD equations:
\begin{enumerate}
  \item For $\alpha(r,\theta,t)$-parameter originated by the CME and given by
  Eq.~(\ref{master}).
  \item For the azimuthal potential $A_{\varphi}(r,\theta,t)$.
  \item For the toroidal magnetic field $B_{\varphi}(r,\theta, t)$.
\end{enumerate}
The differential equations for $A_{\varphi}$ and $B_{\varphi}$ were derived in Ref.~\cite{DvoSem19}.

From standard equation for the axisymmetric azimuthal potential $A_{\varphi}$,
\begin{equation}
  \frac{\partial A_{\varphi}}{\partial t}= \frac{1}{\sigma}
  \left(
    \frac{1}{r}\frac{\partial^2(rA_{\varphi})}{\partial r^2} + \frac{1}{r^2}\frac{\partial}{\partial \theta}  
    \left[
      \frac{1}{\sin \theta}\frac{\partial (\sin \theta A_{\varphi})}{\partial \theta}
    \right]
  \right) +
  \alpha B_{\varphi},
\end{equation}
multiplied by the diffusion time $t_\mathrm{diff}=\sigma R_s^2$, one obtains
\begin{equation}\label{Apotential}
\frac{\partial A}{\partial \tau}=\left[- \mu^2A + \frac{\partial^2A}{\partial \theta^2} + \cot\theta\frac{\partial A}{\partial \theta} - \frac{A}{\sin^2\theta}\right] 
+ \left(\frac{2.55\times 10^{19}}{2}\right)\alpha B, 
\end{equation}
where the dimensionless toroidal magnetic field is normalized on $\rm MeV^2=5\times 10^{13}\,{\rm G}$, i.e. $B=B_{\varphi}/{\rm MeV}^2$. The potential $A_{\varphi}$ is normalized on ${R}_s{\rm MeV}^2$, giving the dimensionless $A=A_{\varphi}/{R}_s{\rm MeV}^2$. At the distance $R_s=150\,\mathrm{km}$ from NS center, the factor $2.55\times 10^{19}/2$ stems from the product $\sigma R_s=2.55\times 10^{19}/2$. 

The standard induction (Faraday) equation for the axisymmetric toroidal field,\footnote{We substitute $r=R_s$ for $\alpha (R_s,\theta,t)$ in a thin layer $\Delta_\mathrm{front}$ within shock front. Thus $\partial_r\alpha=0$ while $\partial_{\theta}\alpha\neq 0$. The other radial derivatives are parametrized in a thin layer as $R_s\times\partial_r(A,B)\to \mathrm{i}\mu (A,B)$, following Parker's suggestion~\cite{Parker} in the standard MHD dynamo.}
\begin{align}
\frac{\partial B_{\varphi}}{\partial t}= &
\frac{1}{\sigma} \left(\frac{1}{r}\frac{\partial^2(rB_{\varphi})}{\partial r^2} + \frac{1}{r^2}\frac{\partial}{\partial \theta}\left[\frac{1}{\sin \theta}\frac{\partial (\sin \theta B_{\varphi})}{\partial \theta}\right]\right)- \frac{1}{r}\frac{\partial}{\partial r}\left[\alpha\frac{\partial}{\partial r}(rA_{\varphi})\right]
\notag
\\
& - \frac{1}{r}\frac{\partial}{\partial \theta}\left[\frac{\alpha}{r\sin\theta}\frac{\partial}{\partial \theta}(\sin \theta A_{\varphi})\right],
\end{align}
being multiplied by $t_\mathrm{diff}=\sigma R_s^2$, takes the following form,
\begin{align}\label{Faraday}
    \frac{\partial B}{\partial \tau}= &
    \left[ -\mu^2B + \frac{\partial^2B}{\partial \theta^2} + \cot\theta \frac{\partial B}{\partial \theta} - \frac{B}{\sin^2\theta}\right]-\left(\frac{2.55\times 10^{19}}{2}\right)
\nonumber
\\
&\times
\left[\frac{\partial \alpha}{\partial \theta}\left(\frac{\partial A}{\partial\theta} + \cot\theta A\right)+ \alpha\left(-\mu^2A+\frac{\partial^2A}{\partial \theta^2} + \cot\theta \frac{\partial A}{\partial \theta} - \frac{A}{\sin^2\theta}\right)\right],
\end{align}
where we account for the coordinates dependence of $\alpha$.

\subsection{Low Fourier mode approximation\label{low}}

In a thin layer $\Delta_\mathrm{front}$ at $r=R_s$, we use the low mode approximation for all dimensionless functions above~\cite{Sokoloff}:
\begin{eqnarray}\label{lowmode}
&&\alpha(t,\theta)=\alpha_1(t)\sin2\theta + \alpha_2(t)\sin 4\theta +...,\nonumber\\
&&A(t, \theta)=a_1(t)\sin \theta + a_2(t)\sin 3\theta +...,\nonumber\\
&&B(t,\theta)=b_1(t)\sin 2\theta + b_2(t)\sin 4\theta +...
\end{eqnarray}
Using the expansions in Eq.~(\ref{lowmode}) and integrating over the colatitude $\theta$, one gets from Eq.~(\ref{master}) that
\begin{align}\label{alpha-evolution}
\dot{\alpha}_1=&\left(\frac{2.86\times 10^{23}}{15\pi (\Delta r)_\mathrm{front}/1\,\text{cm}}\right)(5a_1 - a_2) - 1.36\times 10^{31}\bigg\{\alpha_1 + \alpha_1\frac{2}{\pi}\int_0^{\pi}d\theta\sin^2 2\theta\frac{{\bf B}^2}{B_0^2}
\nonumber
\\
&
+\alpha_2\frac{2}{\pi}\int_0^{\pi}d\theta\sin 2\theta\sin 4\theta\frac{{\bf B}^2}{B_0^2}\bigg\}, \nonumber\\
\dot{\alpha}_2=&
\left(\frac{5.72\times 10^{23}}{105\pi (\Delta r)_\mathrm{front}/1\,\text{cm}}\right)(7a_1 + 37 a_2) - 1.36\times 10^{31}\bigg\{\alpha_2 + \alpha_2\frac{2}{\pi}\int_0^{\pi}d\theta\sin^2 4\theta\frac{{\bf B}^2}{B_0^2}
\nonumber
\\
&
+\alpha_1\frac{2}{\pi}\int_0^{\pi}d\theta\sin 2\theta\sin 4\theta\frac{{\bf B}^2}{B_0^2}\bigg\}, 
\end{align}
where the ratio ${\bf B}^2/B_0^2$ for ${\bf B}^2=B_{\varphi}^2 + B_{\theta}^2 + B_r^2$ is given by
\begin{align}\label{quenching}
\frac{{\bf B}^2}{B_0^2}=&\left(\frac{1}{1.75}\right)\bigg\{b_1^2\sin^22\theta + 2b_1b_2\sin 2\theta\sin 4\theta + b_2^2\sin^24\theta
\nonumber
\\
&
+\mu^2\left[a_1^2\sin^2\theta + 2a_1a_2\sin\theta\sin3\theta + a_2^2\sin^23\theta\right]
\nonumber
\\
&
+ 4(a_1 + a_2)^2\cos^2\theta + 16a_2^2\cos^23\theta + 16(a_1 + a_2)a_2\cos\theta\cos3\theta\bigg\}.  
\end{align}
Here, in the last line, we substituted $(\partial_{\theta}A + \cot\theta A)^2$ for $B_r^2/{\rm MeV}^4$ using Eq. (\ref{components}) and corresponding linear function from Eq.~(\ref{lowmode}), $\partial_{\theta}A + \cot\theta A=2(a_1 + a_2)\cos\theta + 4a_2\cos 3\theta$, and used the expression for $|B_{\theta}|^2/{\rm MeV}^4$ in the second line. 

The evolution equations for all six functions, $a_{1,2}(t)$, $b_{1,2}(t)$ and $\alpha_{1,2}(t)$, have cumbersome form because of the presence of non-linear terms, thereby they are given explicitly in Appendix~\ref{add}. Thus, there are six self-consistent ordinary differential equations: for azimuthal potentials $a_{1,2}$ given by Eq.~(\ref{a1a2}), toroidal field amplitudes $b_{1,2}$ [see Eq.~(\ref{b12full})], and for the helicity parameters $\alpha_{1,2}$ given by Eq.~(\ref{alpha12full}).

\subsubsection{Initial conditions\label{sec:inicond}}

The initial condition for the AMHD magnetic helicity parameter $\alpha(t)=2\alpha_\mathrm{em}\mu_5(t)/\pi\sigma$, which evolves as given in Eq.~(\ref{master}), is chosen relying on a small value of the chiral imbalance, $\mu_5\ll \mu_e\simeq 3\,{\rm MeV}$. We take the initial $\mu_5( 0)\simeq 1\,{\rm MeV}$ at the level of the neutron-proton mass difference. Thus, $\alpha_1(0)= 2.7\times 10^{-4}$, accounting for the conductivity  $\sigma_0= 17\,{\rm MeV}$ at $R_s=150\,\mathrm{km}$ and the temperature $T=10^{10}\,\text{K}$ near the shock front in a nascent NS. We choose also $\alpha_2(0)=\alpha_1(0)/10$ meaning for that a decrease of helicity amplitudes in the low mode approximation.

To get the initial conditions for Eqs.~(\ref{Apotential}) and~(\ref{Faraday}), we equate the initial normalized components \footnote{The dimensionless amplitudes $a_{1,2}$ come from the decomposition in Eq. (\ref{lowmode}) for the dimensionless $A=A_{\varphi}/R_s {\rm MeV}^2$, see definition below Eq. (\ref{Apotential}), where $A_{\varphi}/R_s$ is measured in Gauss, or ${\rm MeV}^2=5\times 10^{13}~{\rm G}$ in units  $\hbar=c=1$. } 
\begin{align}
  \frac{B_{\theta}(R_s,\theta,t=0)}{{\rm MeV}^2}= &
  -\Bigl[a_1(0)\sin\theta + a_2(0)\sin3\theta\Bigr],
  \nonumber
  \\
  \frac{B_r(R_s,\theta,t=0)}{{\rm MeV}^2}= & 2\cos\theta
\Bigl[a_1(0) +a_2(0)(4\cos2\theta - 1)\Bigr],
\end{align}
that come from Eq.~(\ref{components}) in the low mode approximation in Eq.~(\ref{lowmode}), $B_{\theta}(t=0)/{\rm MeV}^2=B_r(t=0)/{\rm MeV}^2=2\times 10^{-4}$.  Then one can find at the same force line of the poloidal field  $B_{p}(t=0)=\sqrt{B_{\theta}^2 + B_r^2}=10^{10}~{\rm G}$, while at different latitudes when substituting corresponding $\theta=0$ for $B_r(t=0)$ where $B_{\theta}(t)=0$  and $\theta=\pi/2$ for $B_{\theta}(t=0)$ where $B_r(t)=0$, the following algebraic system:
\begin{equation}\label{a_initial}
  a_2(0) - a_1(0)=2\times 10^{-4},
  \quad
  a_1 + 3a_2(0)=10^{-4}.
\end{equation}
The initial amplitudes resulting from Eq.~(\ref{a_initial}) have opposite signs, $a_1(0)=-1.25\times 10^{-4}$, $a_2(0)=+ 7.5\times 10^{-5
}$.  We choose the same initial condition for the azimuthal components at $r\simeq R_\mathrm{s}$, $b_{1,2}(0)=2\times 10^{-4}$, corresponding, at the beginning, to the toroidal field at $R_s$, $B(R_s,\theta, t=0)=(\sin 2\theta + \sin 4\theta)\times 10^{10}\,{\rm G}$ in the low mode approximation in Eq.~(\ref{lowmode}).

\section{Results\label{sec:RESULTS}}

\begin{figure}
  \centering
  \subfigure[]
  {\label{1a}
  \includegraphics[scale=.35]{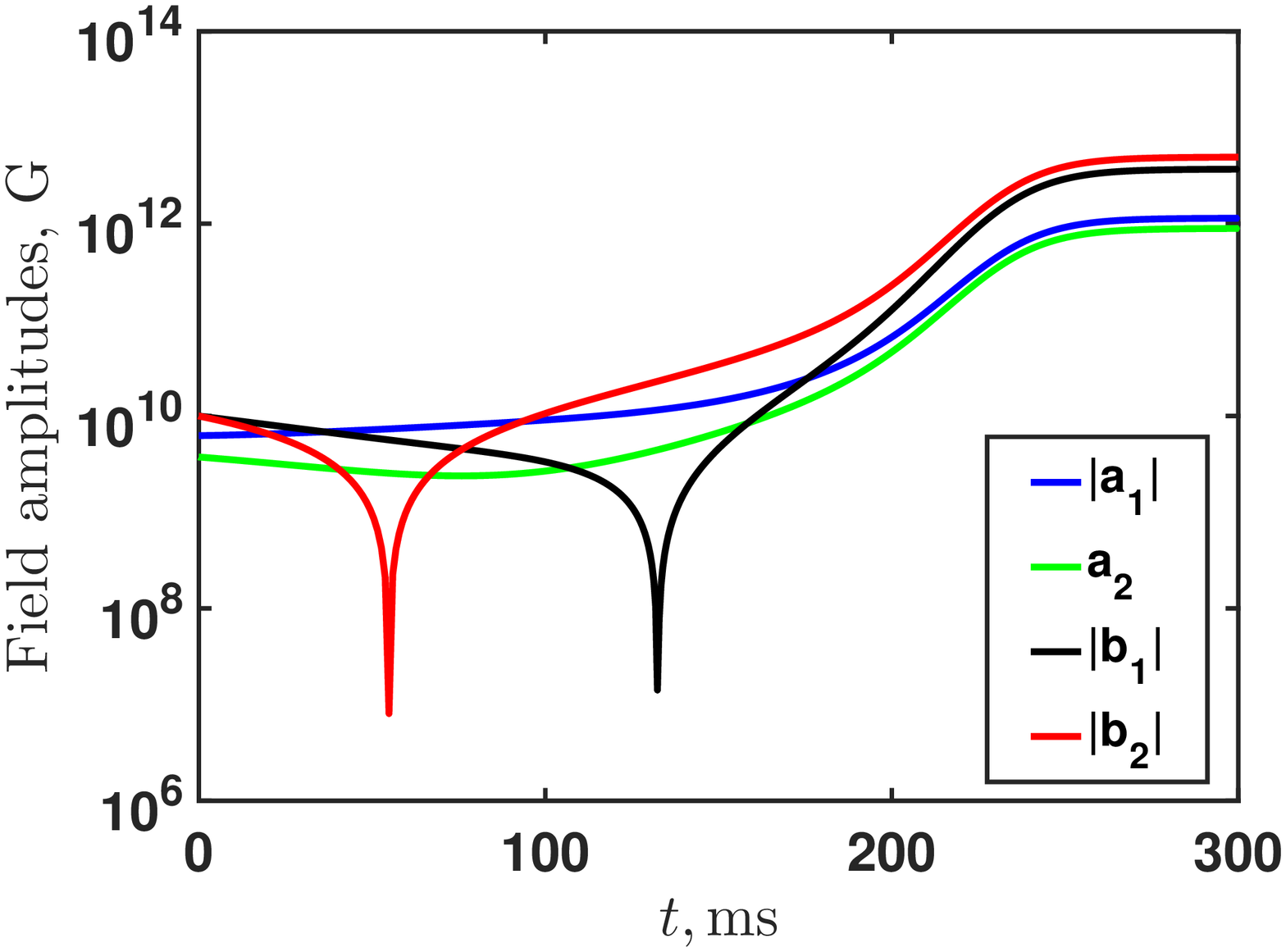}}
  \hskip-.6cm
  \subfigure[]
  {\label{1b}
  \includegraphics[scale=.35]{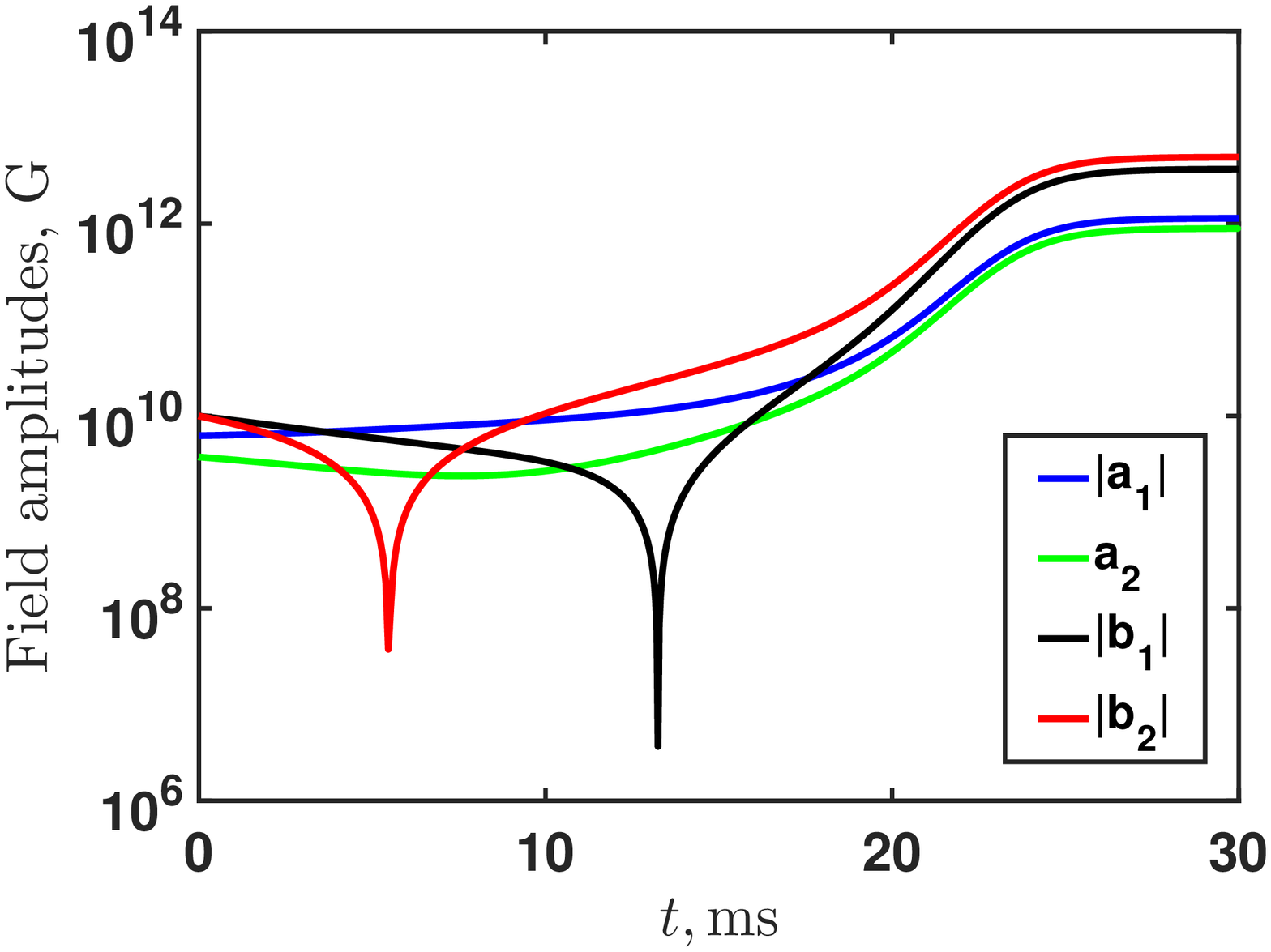}}
  \\
  \subfigure[]
  {\label{1c}
  \includegraphics[scale=.35]{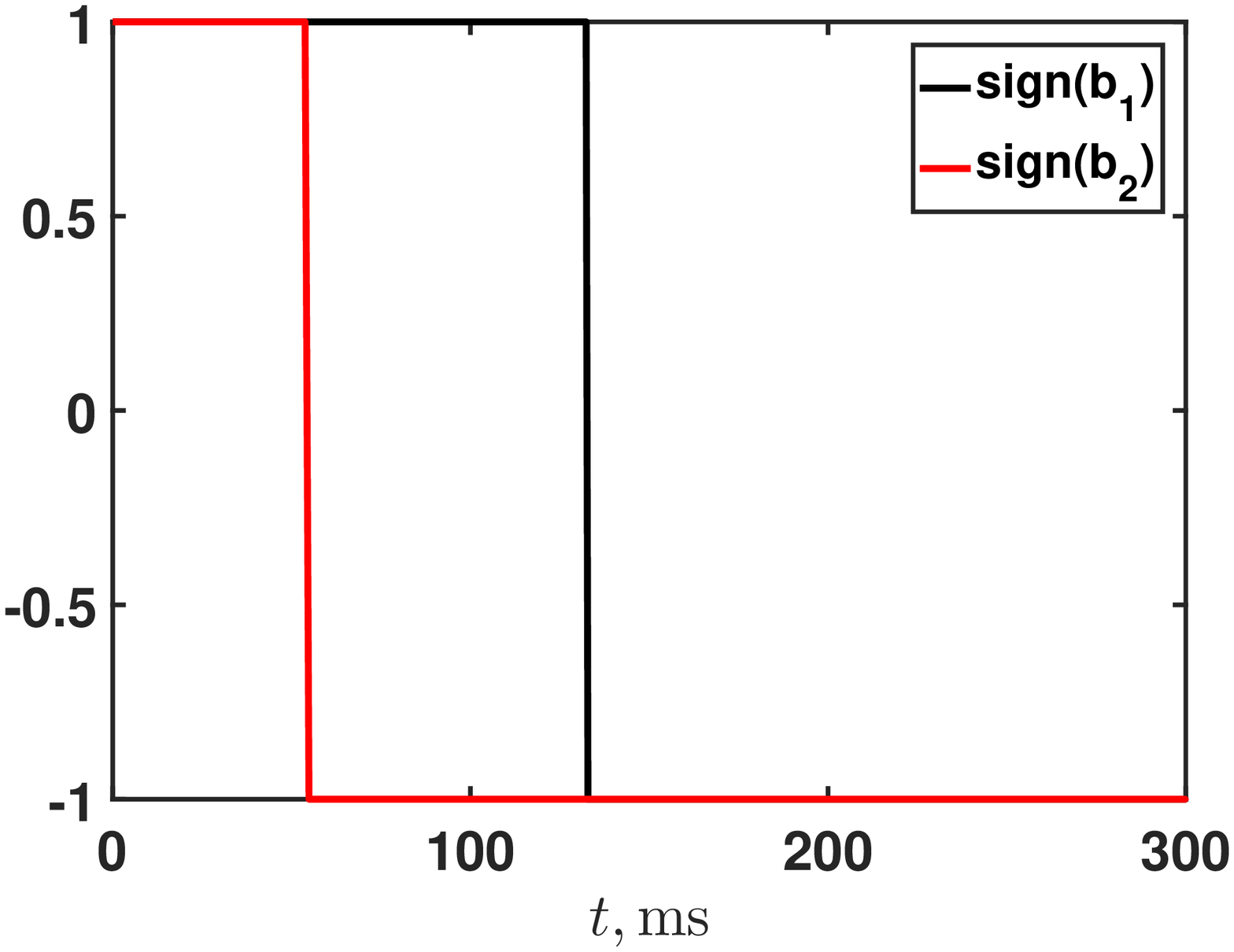}}
  \hskip-.6cm
  \subfigure[]
  {\label{1d}
  \includegraphics[scale=.35]{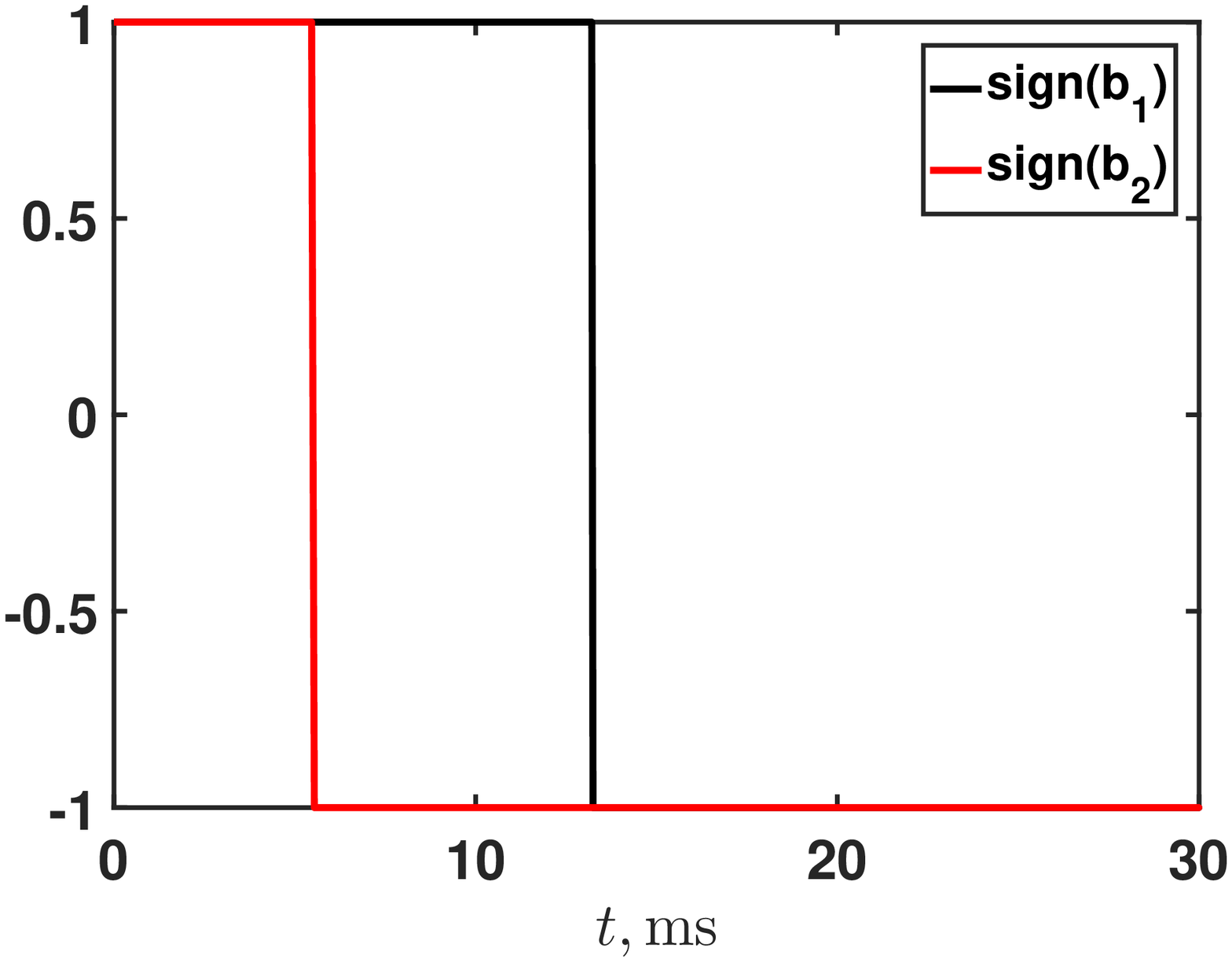}}
  \protect
  \caption{The evolution of the Fourier modes in Eq. (\ref{lowmode}) for the toroidal field 
  $b_{1,2}$ in Eq.~(\ref{Faraday}) and the poloidal field
  $a_{1,2}$ in Eq. (\ref{Apotential}) behind the shock at $r\lesssim R_s=150\,\mathrm{km}$ for 
  the different shock widths.  The magnetic field strength along the vertical axis (measured in Gauss) 
	both for the toroidal and poloidal components:
  (a) $(\Delta r)_\mathrm{front}=10^{-10}\,\text{cm}$;
  (b) $(\Delta r)_\mathrm{front}=10^{-11}\,\text{cm}$.
  The behavior of the signs of $b_{1,2}$ for the different shock widths:
  (c) $(\Delta r)_\mathrm{front}=10^{-10}\,\text{cm}$;
  (d) $(\Delta r)_\mathrm{front}=10^{-11}\,\text{cm}$.
  \label{fig:fieldcomponents}}
\end{figure}

\begin{figure}
  \centering
  \subfigure[]
  {\label{2a}
  \includegraphics[scale=.35]{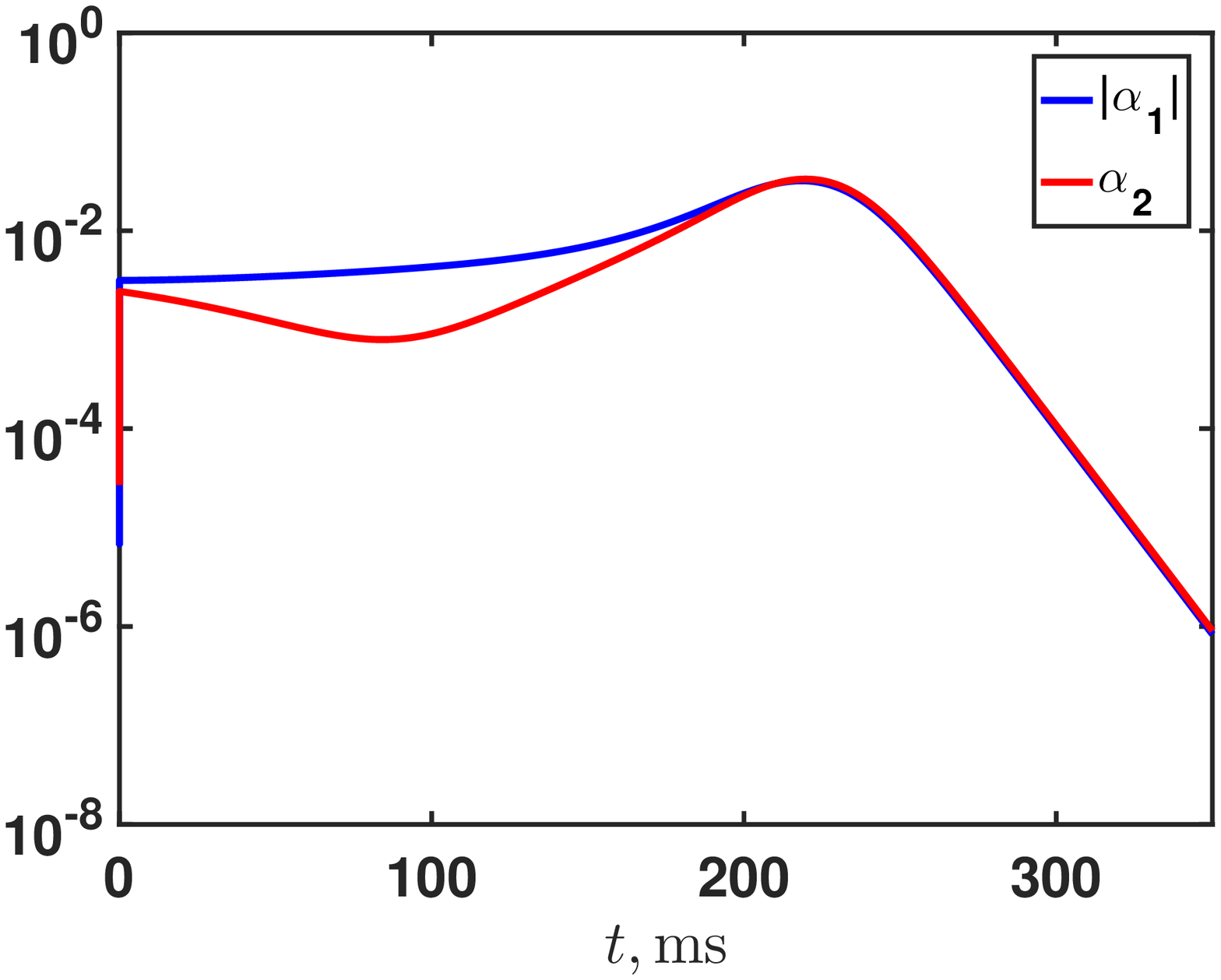}}
  \hskip-.6cm
  \subfigure[]
  {\label{2b}
  \includegraphics[scale=.35]{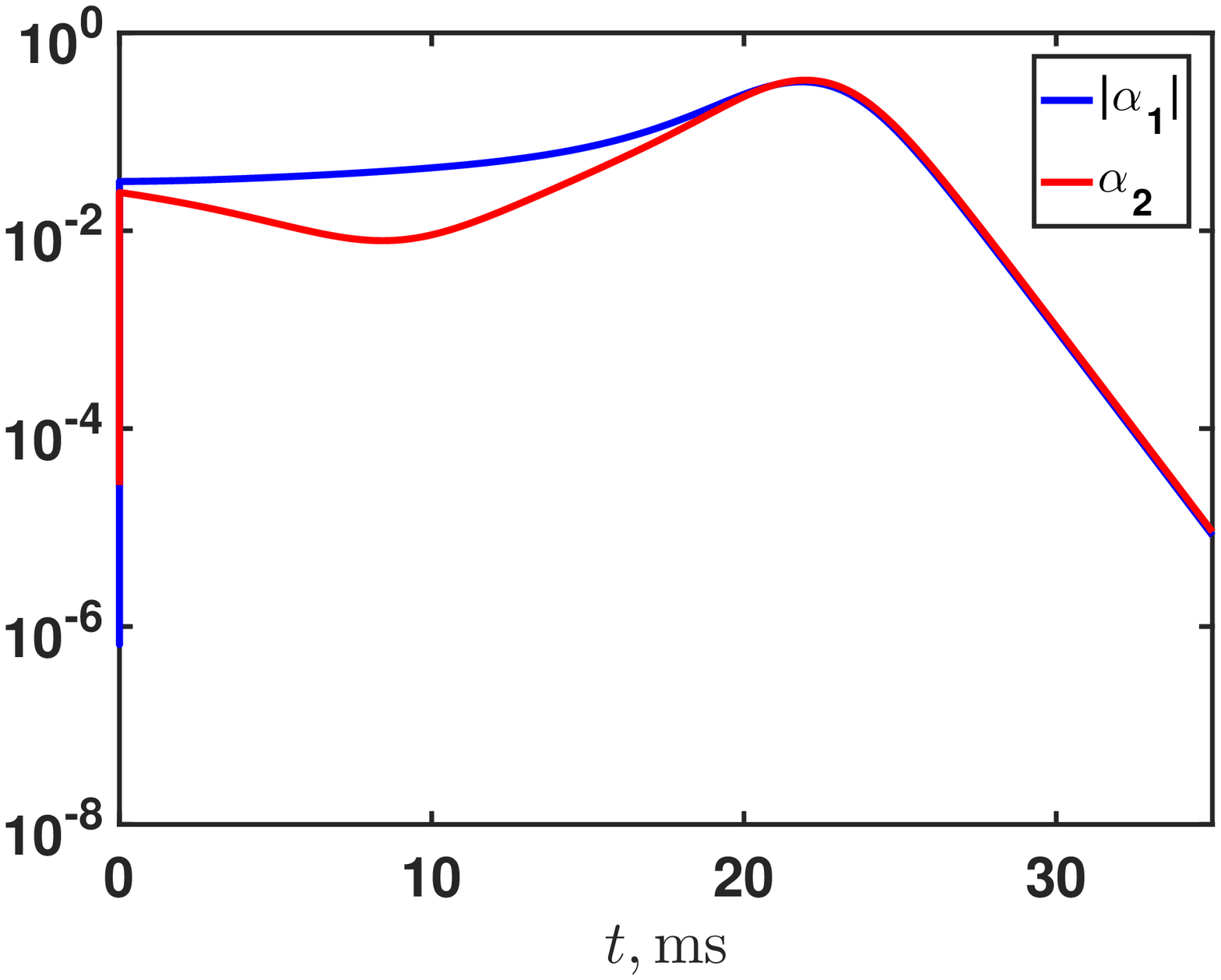}}
  \caption{The evolution of the Fourier modes $\alpha_{1,2}$ in Eq.~(\ref{lowmode}) for the helicity parameter $\alpha$ in Eq.~(\ref{master}) behind the shock at $r\lesssim R_s=150\,\mathrm{km}$: (a) for the shock width $(\Delta r)_\mathrm{front}=10^{-10}\,\text{cm}$, (b) for the shock width $(\Delta r)_\mathrm{front}=10^{-11}\,\text{cm}$.
  \label{fig:alphagrowth}}
\end{figure}

Our results, shown in Figs.~\ref{fig:fieldcomponents} and~\ref{fig:alphagrowth}, are very sensitive to the two parameters: a) the width of the shock front $(\Delta r)_\mathrm{front}$, and b) the time cutoff $t_0$ for CME in Eq. (\ref{cutoff}) that diminishes the helicity $\alpha (t)$ at $t> t_0$  and leads to a saturation of the magnetic field growth at late times after the reversal of the matter inflow to the outflow caused by the shock propagation (forwards only) due to the reheating. We adjust the cutoff time $t_0$ to the corresponding $(\Delta r)_\mathrm{front}$. For that we preset in Eq. (\ref{cutoff}) $\tau_0=3.32\times 10^{-17}$, or $t_0\simeq 200\,\text{ms}$ for $(\Delta r)_\mathrm{front}=10^{-10}\,\text{cm}$, and one order of magnitude less for $(\Delta r)_\mathrm{front}= 10^{-11}\,\text{cm}$: $\tau_0=3.32\times 10^{-18}$, or $t_0\simeq 20\,\text{ms}$. Such a choice is dictated by a shorter shock width for the more intensive shock \cite{LL} which throws out the PNS mantle earlier, consequently interrupting at early time any transfer of magnetized inflow matter onto NS.

There are the two CME outcomes in our AMHD evolution equations. The CME part, given by the term $\sim {\bf B}^2/B_0^2$ in master Eq.~(\ref{master}), arises from the product $\sim {\bf E}\cdot{\bf B}$ in the statistically averaged axial Ward anomaly in Eq.~(\ref{n5evolution}), when we account for the anomalous current ${\bf j}_\mathrm{anom}\sim \mu_5{\bf B}$ in the electric field ${\bf E}$ in Eq.~(\ref{electric}). Such quantum effect occurs negligible at the start since ${\bf B}^2/B_0^2\ll 1$ for the small initial magnetic field $B(t=0)\sim 10^{10}\,{\rm G}$ comparing with the normalization factor $B_0=6.6\times 10^{13}\,{\rm G}$ near the shock $R_s\simeq 150\,\mathrm{km}$, see in Eq.~(\ref{B0}). Then, only at late times $t\gtrsim 10\div 100~ms$, the CME contribution $\sim {\bf B}^2/B_0^2$ becomes effective after a strong growth of the magnetic field components in Fig.~\ref{fig:fieldcomponents} given by the self-consistent {\it non-linear} growth of the $\alpha$-helicity parameter in Fig.~\ref{fig:alphagrowth}. The second CME outcome is stipulated by the presence of the growing helicity $\alpha\sim \mu_5$ shown in Fig.~\ref{fig:alphagrowth} in evolution Eqs.~(\ref{Apotential}) and~(\ref{Faraday}) for the magnetic field components $B_p$ and $B_{\varphi}=B$ correspondingly. 

In Figs~\ref{1a} and~\ref{1b}, one can see the downwardly directed spikes of the toroidal components $b_{1,2}$. This feature can be explained since we use the logarithmic scale in Figs~\ref{1a} and~\ref{1b}. The functions $b_{1,2}$ turn out to change sign in their evolution. In Figs~\ref{1c} and~\ref{1d}, we show the signs of $b_{1,2}$ versus $t$. One can see that the toroidal components become negative exactly at the moments when spikes appear in Figs~\ref{1a} and~\ref{1b}.

\begin{figure}
  \centering
  \subfigure[]
  {\label{3a}
  \includegraphics[scale=.35]{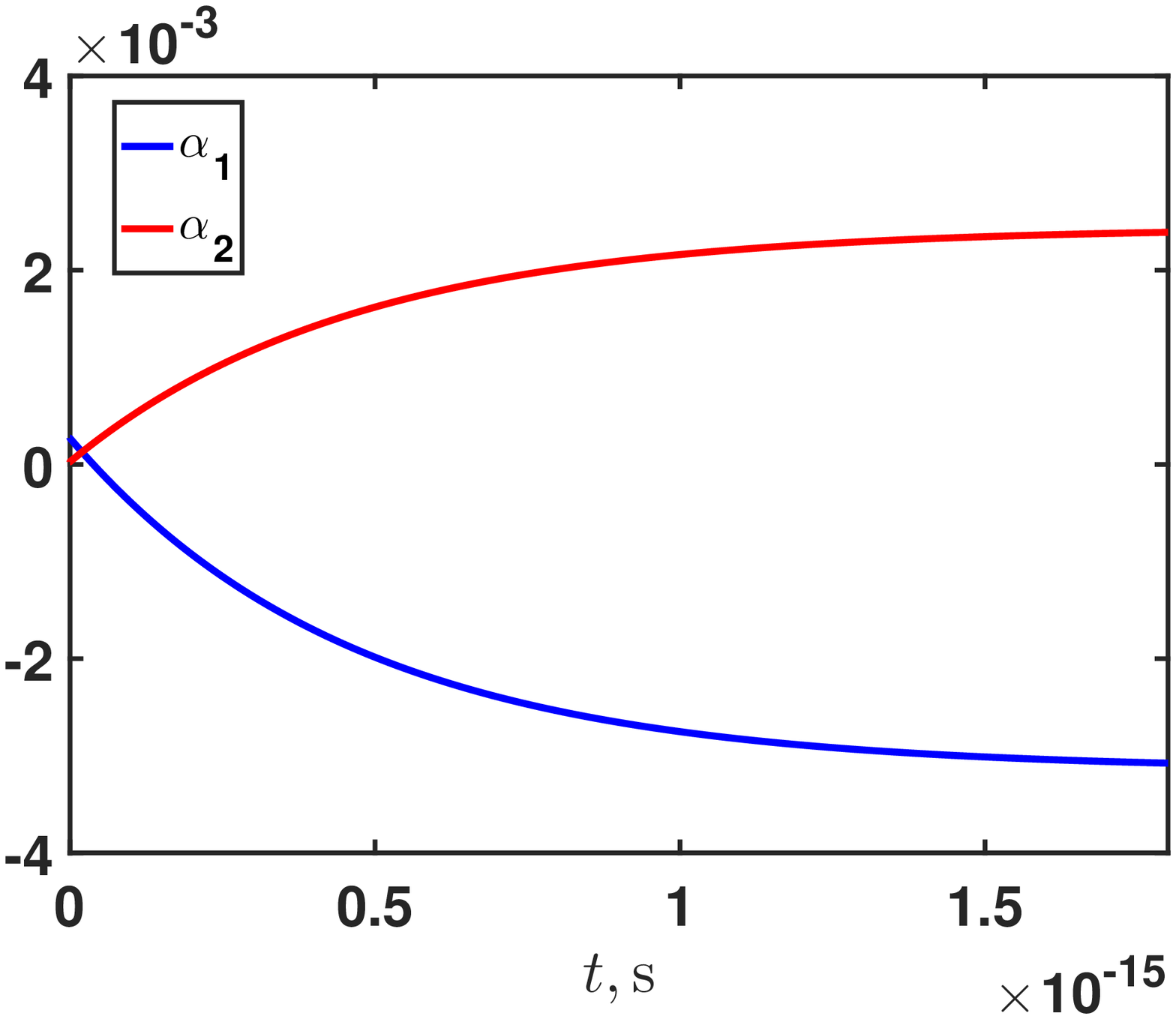}}
  \hskip-.6cm
  \subfigure[]
  {\label{3b}
  \includegraphics[scale=.35]{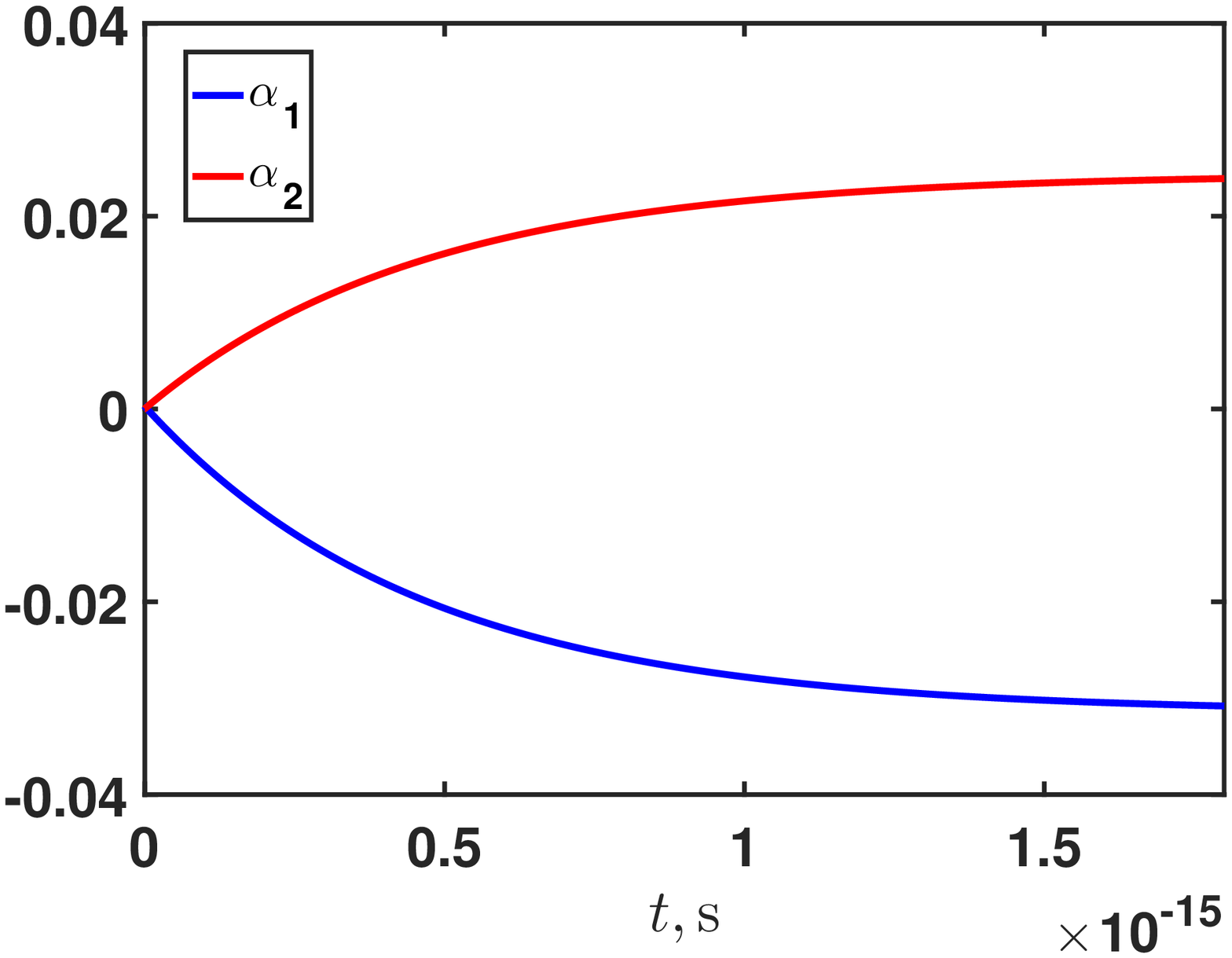}}
	\caption{The spin-flip influence upon helicity components $\alpha_{1,2}$ in Eq.~(\ref{alpha-evolution}) at early times behind the shock at $r\lesssim R_s=150\,\mathrm{km}$: (a) for the shock width $(\Delta r)_\mathrm{front}=10^{-10}\,\text{cm}$, (b) for the shock width $(\Delta r)_\mathrm{front}=10^{-11}\,\text{cm}$.
  \label{fig:alpha12}}
\end{figure}

The influence of the spin-flip $\sim \Gamma_f$ on the helicity $\alpha$ is illustrated in Fig. \ref{fig:alpha12}.
From the start, the shock driver, given by a huge temperature gradient $\sim (\Delta r)_\mathrm{front}^{-1}$ in Eq.~(\ref{alpha-evolution}), leads to the opposite dependencies of the helicity modes $\alpha_{1,2}$ at small times $t\sim 10^{-15}~s$ due to the opposite initial signs of the corresponding coefficients $[5a_1(0) - a_2(0)]<0$ and $[7a_1(0) + 37 a_2(0)]>0$. The negative sign of the spin-flip influence upon $\alpha_{1,2}$ given by the term within braces, $ - 1.36\times 10^{31}\alpha_{1,2}$,  leads to an additional decrease of $\alpha_1$ (see blue line in Fig. \ref{fig:alpha12}), and simultaneously it retards a growth of the mode $\alpha_2$ supported by the positive sign of the corresponding shock driver (see red line in Fig. \ref{fig:alpha12}). Note that, for the shorter shock width $(\Delta r)_\mathrm{front}=10^{-11}\,\text{cm}$, or for a more intensive shock, the helicity modes $\alpha_{1,2}$ rise at early times up to values $\alpha_2$, $|\alpha_1|=(2\div 3)\times 10^{-2}$ shown in Fig. \ref{3b}. This growth is bigger than for the width $(\Delta r)_\mathrm{front}=10^{-10}\,\text{cm}$, for which the modes $\alpha_{1,2}$ reach at the same time $\alpha_2$, $|\alpha_1|=(2\div 3)\times 10^{-3}$, see in Fig. \ref{3a}. While at small times $t\ll t_0$ shown in Fig.~\ref{fig:alpha12} the cutoff factor in Eq.~(\ref{cutoff}) is irrelevant for the evolution of the helicity modes $\alpha_{1,2}$, since $F(t)=1$, it influences strongly at late times diminishing those modes, see in Fig.~\ref{fig:alphagrowth}.

Thus, one can see in Fig.~\ref{fig:fieldcomponents} the growth of both magnetic field components, poloidal $B_p\sim a_{1,2}$ and toroidal $B\sim b_{1,2}$, on three orders of magnitude from a natural initial value $\sim 10^{10}\,{\rm G}$ at the distance $r\gtrsim 150\,\mathrm{km}$ ahead the narrow shock front up to the strong field $B\simeq 10^{13}\,{\rm G}$ behind shock $r\lesssim R_s$. Such amplification is sufficient to get through the following matter inflow onto NS surface the great value $B\sim 2.25\times 10^{15}\,{\rm G}$ in a future magnetar at the NS radius $10\,\mathrm{km}$.

\vskip0.3cm

\section{Discussion\label{sec:DISC}}

In the present work, we suggest the new mechanism for the generation of strong magnetic fields which are produced by the shock after the core bounce  in the SN progenitor of a nascent NS. As the initial condition we assume the presence of the 3D axisymmetric magnetic field in the PNS envelope, ${\bf B}(t, r,\theta)$, having the initial strength $B(t=0)=10^{10}\,{\rm G}$ at the distance $r=R_s=150\,{\rm km}$ from the NS center just before the shock bounce. Such a seed magnetic field arises from the previous contraction of a corresponding pre-SN with the size $r\simeq 10^6\,{\rm km}$ and the magnetic field $B\sim 100\,{\rm G}$ on that time. Then, even without the CME mechanism suggested here, being transferred by matter inflow to the NS surface, such field could reach 
$B\sim  10^{12}\,{\rm G}$ that is a proper value for pulsars. 

The amplification of the seed field $B=10^{10}\,{\rm G}$ in the vicinity of the shock at $r\lesssim R_s$ up to a moderate strong magnetic field $B\sim 10^{13}\,{\rm G}$, see in Fig.~\ref{fig:fieldcomponents} above, does not violate the gas-dynamic description of the shock revival in the model in Ref.~\cite{Janka}. Indeed, for such fields the magnetic energy density is small comparing with the kinetic energy density, $B^2/2\ll \rho v^2/2$. Of course, this condition is fulfilled for the supersonic matter inflow\footnote{The free fall velocity $v\simeq \sqrt{2GM_{\odot}/R_s}=42\times 10^3\,\mathrm{km}/s$ is much greater than the sound velocity $v_s$. The local sound velocity $c_s=\sqrt{4P/3\rho}=5.67\times 10^3~T_{{\rm MeV}}^{1/2}(\mathcal{S}/n)^{1/2}\,\mathrm{km}\cdot\text{s}^{-1}$ is larger in the radiation dominated region behind the shock, $r\lesssim R_s$, since the entropy per baryon $\mathcal{S}/n$ is bigger there developing the strong convection behind the shock due to the convective instability condition, $\partial S/\partial r < 0$. Oppositely to that the sound velocity ahead the shock is less both due to a lower temperature and a smaller entropy.} and the initial magnetic field $B=10^{10}\,{\rm G}= 2\times 10^{-4}\,{\rm MeV}^2$ ahead the shock.
The more important case occurs behind the shock, where the magnetic field rises  up to $B\simeq 10^{13}\,{\rm G}=0.2\,{\rm MeV}^2$ owing to the AMHD dynamo, while fluid velocities are smaller than the local sound velocity due to the deceleration by the shock, $v< c_s$. Nevertheless, for $v\lesssim c_s=10^4\,\mathrm{km}\cdot\text{s}^{-1}$ and $\rho\simeq 10^9\,\text{g}\cdot\text{cm}^{-3}$ in that region, the kinetic energy density  $\rho v^2/2= 2.5\,{\rm MeV}^4$ is much bigger than the magnetic energy density of the saturated magnetic field at the shock $R_s=150\,\mathrm{km}$, $\rho v^2/2\gg B^2/2=0.02\,{\rm MeV}^4$.  Thus, the gas-dynamic description prevails. For the same reason we use the electric conductivity $\sigma$ in Eq.~(\ref{conductivity}) valid for a dense non-magnetized {\it npe} matter~\cite{Shternin} instead of the conductivity tensor $\sigma_{ij}$ for a magneto-active plasma. 

The strong amplified magnetic field $B\simeq 10^{13}\,{\rm G}$, frozen in plasma behind the shock, is transferred by the inflow crossing shock front at $R_s=150\,{\rm km}$ onto NS surface at $r=R_\mathrm{NS}=10\,{\rm km}$, hence reaches there the value $B(R_\mathrm{NS})\simeq 2.3\times 10^{15}\,{\rm G}$ that is typical for magnetars. This amplification proceeds unless the shock being reheated converts the inflow to the outflow throwing the whole mantle. That moment is the end of the CME activity driven by the temperature gradient $\mathrm{d}T/\mathrm{d}r$ at the shock front, the magnetic helicity parameter $\alpha\sim \mu_5$ ceases, $\alpha\to 0$, see in Fig.~\ref{fig:alphagrowth}, and we imitate that dynamic shutdown by the smooth temporal factor in Eq.~(\ref{cutoff}) ahead $\mathrm{d}T/\mathrm{d}r$. In Eq.~(\ref{cutoff}), the cutoff time $t_0$ varies in dependence on a shock intensity and the corresponding shock width, see also comments in Sec.~\ref{scenario}. We obtain numerically $t_0\sim 20\,\text{ms}$ for $(\Delta r)_\mathrm{front}=10^{-11}\,\text{cm}$ and $t_0\sim 200\,\text{ms}$ for $(\Delta r)_\mathrm{front}=10^{-10}\,\text{cm}$, seen as the corresponding moments at the beginning of the saturation for ${\bf B}$ in Fig.~\ref{fig:fieldcomponents}. 

The role of the convection and the advection of matter behind the shock is sub-dominant for the successful SN explosion when the neutrino time reheating is the shortest one, $\tau_\mathrm{heat}< \tau_\mathrm{cv}\sim \tau_\mathrm{ad}$~\cite{Janka2}. This inequality, on one hand, leads to the nearly spherical explosion, see in Fig.~6 in Ref.~\cite{Janka2}.
On the other hand, the fast reheating means the sub-dominant role of turbulent velocities $v\ll c_s$ behind the shock responsible for the origin of the helicity $\alpha=\tau_\mathrm{corr}\langle{\bf v}\cdot(\nabla\times {\bf v}\rangle/3$ in the standard MHD. We see in Fig.~\ref{fig:alphagrowth} that the  $\alpha$-helicity parameter in our AMHD model is competitive to $\alpha\simeq 10^{-3}$ in the $(\alpha-\Omega)$-dynamo model in Ref.~\cite{Thompson:1993hn} based on convective motions which are caused by the convection instability $\mathrm{d}S/\mathrm{d}r<0$ in a nascent NS. Indeed, in the AMHD dynamo model suggested here the helicity $\alpha$ becomes even greater at its maximum, coinciding over time with moments of saturation for ${\bf B}$ and reaches $\alpha\sim 0.02$ for $(\Delta r)_\mathrm{front}=10^{-10}\,\text{cm}$, or $\alpha\sim 0.1$ for $(\Delta r)_\mathrm{front}=10^{-11}\,\text{cm}$, see in Fig.~\ref{fig:alphagrowth}. 

In the magnetorotational model developed with the use of differential rotation, the toroidal magnetic field is growing linearly with time $B_{\varphi}\sim B_r\omega t$. Therefore it can not reach sufficiently strong values with the timescale compared to the accretion timescale of the collapsed core (see, e.g., the model in Ref. \cite{Ardeljan:2004fq} and comments on that
in Ref. \cite{Janka:2002eg}). 

To resume, the new CME mechanism based on the axial Ward anomaly in particle physics [see Eq.~(\ref{pseudovector})] provides the growth of natural seed magnetic fields in a SN progenitor of PNS up to the strong fields $B\sim 2\times 10^{15}\,{\rm G}$ in a nascent NS. The amplification of  the magnetic field driven by the shock in such SN can be effective even without usual shock consequences like the postshock advection or convective motions, being similar to the $\alpha^2$-dynamo known in standard MHD. 

\section*{Acknowledgments}

We are thankful to D.K.~Nadyozhin and L.B.~Leinson for comments on models of supernovae explosion, as well as to M.~Gusakov and A.~Chugunov for comments on the transport properties of dense matter in NS. This work is supported by the government assignment of IZMIRAN. M.~Dvornikov is indebted to the Russian Science Foundation (Grant No.~19-12-00042) for the support. D.~Sokoloff acknowledges RFBR for the support under the Grant No.~18-02-00085.

\appendix

\section{Complete set of AMHD evolution equations in the mean field dynamo for rigid NS rotation\label{add}}

In this appendix, we present the full set of differential equations for the expansion coefficients in Eq.~\eqref{lowmode}.

First, we substitute the low mode series Eq.~(\ref{lowmode}) in Eq.~(\ref{Apotential}). Then, we multiply it by $\sin \theta$ and $\sin 3\theta$, correspondingly, and integrate that over the colatitude angle, $(2/\pi)\int_0^{\pi} d\theta (\dots)$, using the normalization
\begin{equation}
  \frac{2}{\pi}
  \int_0^{\pi}\sin^2(m\theta)\mathrm{d}\theta=1,
  \quad
  m=1,2,
\end{equation}
and the orthogonality
\begin{equation}
  \frac{2}{\pi}
  \int_0^{\pi}\sin (m\theta)\sin (n\theta)\mathrm{d}\theta=0,
  \quad
  m\neq n.
\end{equation}
Finally, we obtain the system of the ordinary non-linear differential equations for the azimuthal potential components $a_{1,2}$,
\begin{align}\label{a1a2}
\dot{a}_{1} =&-\left[(\mu^{2}+2)a_{1}+2a_{2}\right]+ 2.55\left(\frac{16\times 10^{19}}{315\pi}\right)\left(21\alpha_{1}b_{1}-6(\alpha_{1}b_{2}+\alpha_{2}b_{1})+ 20\alpha_{2}b_{2}\right),
\nonumber
\\
\dot{a}_{2} =&-(\mu^{2}+12)a_{2}+ 2.55\left(\frac{16\times 10^{19}}{\pi}\right)\left(\frac{4}{165}\alpha_{2}b_{2}+\frac{2}{45}(\alpha_{1}b_{2}+\alpha_{2}b_{1})+\frac{\alpha_{1}b_{1}}{21}\right).
\end{align}
Then, we again use the low mode approximation in the Faraday Eq.~(\ref{Faraday}) and multiply it by $\sin 2\theta$ and $\sin 4\theta$, correspondingly. After the integration it over the colatitude angle $(2/\pi)\smallint_0^{\pi} \mathrm{d}\theta (\dots)$, we obtain the system of the differential equations for the toroidal magnetic field amplitudes $b_{1,2}$,
\begin{align}\label{b12full}
\dot{b}_{1}=  & -\left[(\mu^{2}+6)b_{1}+4b_{2}\right]
\nonumber
\\
& + 2.55\left(\frac{16\times 10^{19}}{315\pi}\right)\Bigl[\mu^2\Bigl(21a_{1}\alpha_{1}-6a_{1}\alpha_{2}+15a_{2}\alpha_{1}+14a_{2}\alpha_{2}\Bigr)
\nonumber
\\
& + 21a_{1}\alpha_{1}+ 66a_{1}\alpha_{2}+ 111a_{2}\alpha_{1}+ 38a_{2}\alpha_{2}\Bigr],\nonumber\\
\dot{b}_{2}= & -(\mu^{2}+20)b_{2}- 2.55\left(\frac{32\times 10^{19}}{3465\pi}\right)\Bigl[\mu^2\Bigl(33a_{1}\alpha_{1}- 110a_{1}\alpha_{2}-77a_{2}\alpha_{1} -42a_{2}\alpha_{2}\Bigr)
\nonumber
\\
& + 429a_{1}\alpha_{1}-110a_{1}\alpha_{2}-649a_{2}\alpha_{1}-362a_{2}\alpha_{2}\Bigr].
\end{align}
Finally, the full set of AMHD evolution equations is completed by the two evolution equations for $\alpha_{1,2}(t)$ helicity amplitudes resulting from Eq.~(\ref{alpha-evolution}),
\begin{align}\label{alpha12full}
\dot{\alpha}_{1}= &  \frac{2.86\times10^{23}}{15\pi (\Delta r)_\mathrm{front}}(5a_{1}-a_{2})-1.36\times10^{31}\alpha_{1}
\nonumber
\\
 & -1.77\times10^{25}\Bigl(\mu^{2}\Bigl[2a_{1}^{2}\alpha_{1}- a_{1}^{2}\alpha_{2}+2a_{1}a_{2}\alpha_{1}+2a_{1}a_{2}\alpha_{2}
\nonumber
\\
\displaybreak[2]
 & +2a_{2}^{2}\alpha_{1}+ a_{2}^{2}\alpha_{2}\Bigr]+8a_{1}^{2}\alpha_{1}+4a_{1}^{2}\alpha_{2}+24a_{1}a_{2}\alpha_{2}+24a_{2}^{2}\alpha_{1}
\nonumber
\\
& +4a_{2}^{2}\alpha_{2}+3\alpha_{1}b_{1}^{2}+2\alpha_{1}b_{2}^{2}+4\alpha_{2}b_{1}b_{2}\Bigr),
\nonumber
\\
\displaybreak[2]
\dot{\alpha}_{2}= & \frac{5.72\times10^{23}}{105\pi (\Delta r)_\mathrm{front}}\left(7a_{1} + 37a_{2}\right)-1.36\times10^{31}\alpha_{2}
\nonumber
\\
\displaybreak[2]
 & +1.77\times10^{25}\Bigl(\mu^{2}\Bigl[a_{1}^{2}\alpha_{1}-2a_{1}^{2}\alpha_{2}-2a_{1}a_{2}\alpha_{1}-a_{2}^{2}\alpha_{1}-2a_{2}^{2}\alpha_{2}\Bigr]
\nonumber
\\
\displaybreak[2]
& -4a_{1}^{2}\alpha_{1}-8a_{1}^{2}\alpha_{2}-24a_{1}a_{2}\alpha_{1}-16a_{1}a_{2}\alpha_{2}-4a_{2}^{2}\alpha_{1}-40a_{2}^{2}\alpha_{2}
\nonumber
\\
& -2\alpha_{2}b_{1}^{2}-3\alpha_{2}b_{2}^{2}-4\alpha_{1}b_{1}b_{2}\Bigr),
\end{align}
where we use the factor ${\bf B}^2/{\rm B}^2$ given in Eq.~(\ref{quenching}). We put also $\mu=0$ in our numerical calculations since Parker's suggestion for the radial derivatives in dynamo~\cite{Parker}, $R_s\partial_r (A,B)\to -\mathrm{i}\mu (A,B)$, has no sense for the superfine shock width $(\Delta r)_\mathrm{front}$ at $r=R_s$.

%

\end{document}